\documentclass[twocolumn]{aastex631}

\usepackage{amsmath}
\usepackage{makecell}
\usepackage{tabularx}
\usepackage{hyperref}

\begin{document}

\title{Planetary Rhythms: Synchronous Circulation on Variably Irradiated Asynchronous Planets}

\author[0000-0002-7179-8254]{Deepayan Banik}
\affiliation{Department of Physics, University of Toronto, 60 St George Street, Toronto, Ontario, M5S 1A7, Canada}

\begin{abstract}

Tidal locking of planets to their host stars results in an atmospheric circulation with a hotspot fixed to the frame of reference of the planet. On the other hand, asynchronously rotating planets feature moving hotspots either lagging or leading the corresponding substellar point as it translates along the surface. We show that a planet falling in the latter category could mimic the circulation of tidally synchronous planets under the influence of time-varying instellation, possibly provided by pulsating or multiple star systems. This happens when the planet's diurnal period is in resonance with the period of instellation variation, leading to a planet-frame-fixed hotspot. Slight differences in the above periods lead to East-West or West-East creeping hotspots with a period significantly longer than both. The rate of hotspot motion is given by the difference between the diurnal and instellation variation rates, similar to the lower envelope frequency of beat patterns formed by two superposed waves in linear wave theory. We call this phenomenon `beating'. A combination of the radiative, rotational, wave propagation, and drag timescales establishes dynamical constraints on beating. Based on this we classify a set of Kepler and TESS circumbinary planets with two candidates exhibiting climatic departures from the no-variation scenario. In general, hotter and faster-spinning planets are more susceptible to climatic departures. Beating, if it occurs, may additionally create optimistic extensions of habitable zones for corresponding systems. 

\end{abstract}



\section{Introduction} \label{sec:intro}


In recent popular science fiction literature, the Three Body Problem, Chinese novelist Cixin Li portrays a hypothetical variably irradiated world that exhibits chaotic weather patterns, quite the opposite of our experience of the Earth. Notwithstanding the extreme scenario, the major takeaway is that a planet's climate is strongly influenced by the type of radiation it receives. At first order, this is primarily linked to its rotation and orbital periods. When equal, the planets have a permanent day side and are called tidally synchronous, otherwise, the day and night coverage shifts continuously resulting in asynchronous planets like the Earth. Spatio-temporal variation of irradiation may also arise from the planet's precession, orbital eccentricity, or axial tilt leading to well-known Milankovitch cycles \citep{deitrick2018exo}. All these effects have been extensively studied in the literature \citep{pierrehumbert2019atmospheric, penn2017thermal, ohno2019atmospheres}. 

The discovery of planets around binary stars \citep{borucki2010kepler} added another source of variability to the list, the circumbinary ``gyration'' effect --- the planet-relative orbital motion of binary stars \citep{haqq2019constraining}. Additionally, stars like RR Lyrae variables, Delta Scuti variables, or short-period Cepheids \citep{dambis2013rr,handler2009delta,csornyei2022study} are known to feature stellar pulsations, another potential cause of inherent variability. Planets around such systems are envisaged to feature interesting phenomena due to resonance effects \citep{forgan2014assessing}, yet, much is left to be understood so far as climate modeling is concerned. 

\begin{figure*}
    \centering
\includegraphics[width=\linewidth]{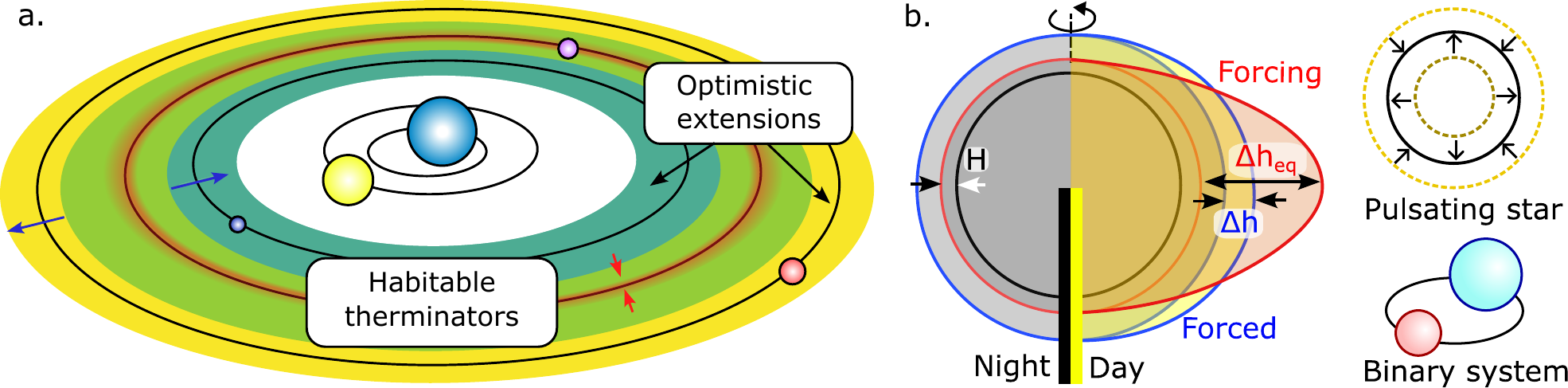}
    \caption{a. Schematic of a circumbinary habitable zone (green) with regions of therminator habitability inside it (red shade) and, possible optimistic extensions within (blue) and beyond (yellow) the inner and outer edges, respectively, in the case of beating planets. b. Schematic of the shallow water model used in the simulations with different relevant heights serving as a proxy for atmospheric temperature affected by a periodically varying instellation source. The red line corresponds to the radiatively forced profile, while the blue represents the steady-state equilibrium height. }
    \label{fig:fig1}
\end{figure*}

Previous atmospheric studies of variably irradiated planets (VIP) have used 1-D Energy balance models (EBM) owing to their simplicity and cost-effectiveness \citep{forgan2014assessing,may2016examining,haqq2019constraining,yadavalli2020effects}. However, EBMs apply well to planets with a zonally averaged circulation, thereby precluding periodic variations in the local climate engendered by instellation variation. This has likely led to the conclusion that the atmospheric dynamics of gaseous circumbinary planets (CBP) differ little from the equivalent single-star case \citep{may2016examining}. Additionally, global climate models (GCM) for circumbinary Earth-like planets have shown only minor effects on the surface climate  \citep{popp2017climate,wolf2020resilience}. Moreover, the computational expense of GCMs allows for only a small survey of the parameter space excluding regimes of potential climate variability and extremes. Consequently, concerns of habitability are limited to these cases \citep{komacek2020stable}.

Given the above, we ask the following questions: Can time-varying instellation alter the atmospheric circulation of planets significantly? If yes, under what conditions, and to what extent? What is the nature of the resulting climate variability, and how does it differ from the no-variation scenario?

In this paper, we present a first survey of variably irradiated planetary atmospheres using a shallow-water model (SWM). Principally, we demonstrate that the atmospheric circulation of Earth-like and gaseous VIPs could differ significantly from the no-variation case, contrary to previous research. This happens in a specific regime of parameter space where the diurnal and irradiation variation frequencies are close or synced. In such cases, an asynchronous rotator may \textit{mimic} the circulation pattern of a tidally locked planet with a {\color{black} permanent planet-frame-fixed global temperature maxima or hotspot}, resulting in local temperature extremes. We define a new term called the `therminator' to demarcate regions of thermal opposites. Minor deviations from the aforementioned synchronization lead to long-term climate variability, with the therminators slowly creeping across the planet surface. The gradual motion of the polarized global climate occurs at a rate given by the difference of the diurnal and irradiation variation frequencies, much like the lower envelope frequency of beat patterns formed by two linearly superposed sound waves. We refer to this dynamical phenomenon as `beating'. 

The current work points to multiple directions of further research. Fig. \ref{fig:fig1}a shows the traditional circumbinary habitable zone (HZ) in green, where a beating planet might carve out a relatively uninhabitable region due to the creation of thermal extremes on opposite sides of the planet. The therminators of such a planet could however be habitable \citep{lobo2023terminator}. On the other hand, beating might lead to optimistic extensions of HZs on both the inner and outer edges, with habitable cold and hot sides, respectively. Thus, the potential for diverse climates on VIPs opens up a new area of exploration for atmospheric dynamicists and habitability experts alike. Additionally, the synchronization of planetary diurnal/rotation periods with the instellation variation period of their host stellar systems requires investigation. A potential role may be played by thermal tides, already known to cause Earth-like planets around single low-mass stars to rotate asynchronously \citep{leconte2015asynchronous}. 

In \S \ref{sec:2} we briefly describe our two-layer spherical shallow-water mathematical model. In \S \ref{sec:3} we discuss the results: the beating phenomena and therminators are defined as relevant to the work. Additionally, we provide a first-order analytical reasoning for the occurrence of beating. We also obtain limits on known dynamical timescales to map out a region of parameter space where beating might occur. In \S \ref{sec:4} we evaluate specific Kepler and TESS CBPs based on the above criteria and make general comments on CBP habitability. Finally, we conclude in \S \ref{sec:5}.

\section{Mathematical model}\label{sec:2}

\subsection{Model setup}

We use an idealized, spherical (latitude($\phi$) - longitude($\lambda$) grid), two-layer shallow-water model \citep{vallis2017atmospheric} to simulate the atmosphere of a non-eccentric-tilted planet. Such models have been extensively used to study the atmospheric dynamics of exoplanets in multiple regimes of rotation, radiation, drag, etc. \citep{perez2013atmospheric, ohno2019atmospheres, penn2017thermal}. Being 2D, SWMs capture more physical aspects of atmospheric modeling than 1D EBMs, and are also less expensive than 3D GCMs, making them perfect tools for exploration of dynamical phenomena, the prime focus of our work. 


The model features two constant density layers: an active upper layer of height $h(\lambda,\phi,t)$ (blue line, Fig. \ref{fig:fig1}b) overlying an inert layer of infinite depth (black line). The governing equations of our model involve the depth-averaged in-plane momentum and mass balances as follows \citep{perez2013atmospheric}, 
\begin{equation}
    \frac{\partial \mathbf{u}}{\partial t} + (\mathbf{u} \cdot \nabla) \mathbf{u} + f\mathbf{k} \times \mathbf{u} + g\nabla h = \textbf{R} - \frac{\mathbf{u}}{\tau_{\text{drag}}} \label{eq:1}
    \end{equation}
    \begin{equation}
    \frac{\partial h}{\partial t} + \nabla \cdot (h\mathbf{u}) = \frac{h_{\text{eq}}(\lambda,\phi,t)-h}{\tau_{\text{rad}}} \equiv \textit{Q} \label{eq:2}
\end{equation}
\begin{equation}
    \textbf{R} = -\frac{\mathbf{u}}{h} \cdot\frac{Q+|Q|}{2} \label{eq:3}
\end{equation}
where $\mathbf{u}(\lambda,\phi,t)$ is the horizontal velocity, $f=2 \Omega \sin\phi$ is the Coriolis parameter, $\mathbf{k}$ is the unit vector along the axis of planetary rotation, $g$ is the gravitational acceleration, $h$ is the fluid layer thickness, and $h_{\text{eq}}$ represents the equilibrium height profile. $\tau_{\text{drag}}$ is the Rayleigh drag timescale incorporating flow resistive effects of Lorentz forces \citep{perna2010magnetic}, turbulent mixing \citep{li2010circulation}, gravity wave breaking \citep{watkins2010gravity} or bottom friction \citep{koll2018atmospheric}, whichever applicable. The Newtonian radiative cooling timescale $\tau_{\text{rad}}$, is dependent on multiple things like the orbital separation, stellar properties, temperature-pressure profile of the planetary atmosphere and its composition \citep{ohno2019atmospheres}.

The height of fluid in the top layer is used as a representative of temperature. Consequently, the radiative forcing is described by $h_{\text{eq}}$, a profile to which the height field is relaxed over the radiative timescale described above (Fig. \ref{fig:fig1}b). We implement stellar variability as a periodic fluctuation about a mean according to the following expression,
\begin{equation}
     h_{\rm{eq}}(\lambda,\phi,t) = \rm{H} + \Delta h_{eq}\cdot\frac{\mathcal{R} + |\mathcal{R}|}{2} \left(1 + \bar{\rm{f}} \cos \omega t \right) \label{eq:4}
\end{equation}
where, 
\begin{equation}
    \mathcal{R} = \cos (\lambda - \lambda_0(t)) \cos\phi, \hspace{1mm} \text{and} \hspace{1mm} \lambda_0(t) = \Omega t.
    \label{eq:5}
\end{equation}
Here, $\lambda_0(t)$ present represents the present longitude of the substellar point as applicable to asynchronous planets \citep{penn2017thermal,ohno2019atmospheres}. This is valid under the assumption that the orbital period is sufficiently long, such that the rotational frequency dominates the length of the day, and hence $\lambda_0(t)$. Henceforth, the terms `rotational' and `diurnal' will thus be used interchangeably. $\rm{H}$ is the mean nightside thickness and $\Delta \rm{h_{eq}}$ is the day-night height contrast in radiative equilibrium. The semiamplitude and frequency of variation of instellation are $\bar{\rm{f}}$ and $\omega$, respectively. The radiated planetary hemisphere receives maximum instellation at the substellar point dying off as cosines along both $\lambda$ and $\phi$ away from it. The other hemisphere is forced to a uniform height equal to the initial nightside thickness H, implying that no radiation is received. The term $\rm{\textbf{R}}$ represents mass and momentum transfer between the two layers and is important for several reasons including conservation properties \citep{showman2010matsuno}, reproduction of equatorial superrotation \citep{showman2011equatorial}, and, convergence to a single statistical equilibrium state irrespective of initial conditions \citep{liu2013atmospheric}. Finally, the factor $(\mathcal{R} + |\mathcal{R}|)/2$ makes the implementation of hemispheric irradiation computationally convenient \citep{dobrovolskis2009insolation}.

\begin{table}
\centering
\hspace{-1.5cm}
\begin{tabular}{cc|cc}
\hline
Parameter & Value & Parameter & Value \\
\hline
\hline
$\tau_{\text{rad}}$ & 5 days & $\tau_{\text{drag}}$ & 5 days \\
$P_{\text{rot}}$ & 1 days & Radius & 6371 km \\
$g$ & 10 ms$^{-2}$ & H & 100 m \\
$\bar{\text{f}}$ & 0.5 \\
\end{tabular}
\caption{Parameter values of the Earth-like control simulation taken from \cite{ohno2019atmospheres}.}
\label{tab:contr}
\end{table}

\subsection{Numerical scheme}

We use Dedalus3 \citep{burns2020dedalus} to numerically solve equations (\ref{eq:1}) and (\ref{eq:2}). Dedalus is a pseudo-spectral code used to solve initial, boundary, or eigen-value problems with the flexibility of specifying the equations symbolically. A sphere basis is used to discretize our 2-D domain. Post discretization the system is evolved through the SBDF2 timestepper. The simulation is resolved in 32 longitudes and 16 latitudes, which is low but sufficient to capture the central physics phenomena discussed ahead. The typical timestep size is 0.1 hours and the total simulation time is usually the minimum of twice the longest relevant timescale or 500 days.  

\subsection{Parameter choices}

We choose the parameters for an Earth-like planet \citep{penn2017thermal} i.e. a radius of 6371 km, ${g\text{H}}=1000 \text{ m}^2\text{s}^{-2}$ and $\tau_{\text{rad}}=\tau_{\text{drag}}=5$ days, and $\bar{\rm{f}}$ is chosen to be 0.5 based on the maximum estimate for joint-binary-star-flux semiamplitude in \cite{haqq2019constraining}; see Table \ref{tab:contr}. The numerical model is set up with initial conditions of zero velocity and uniform height equivalent to the nightside height {\color{black} H (=100 m)}. The initial location of the substellar point is at 180$^\circ$. The system is then progressively iterated over time until it reaches a statistical steady state. The term `statistical' is crucial for two reasons: one, the moving substellar point keeps shifting the location of forcing throughout the simulation, and, two, the instellation varies with time. Thus a true steady state may not be reached as long as the simulation runs. In that case, the saturation of area-averaged kinetic energy or average height is used to determine the statistical steady state. 

\begin{figure}
    \centering
\includegraphics[width=\linewidth]{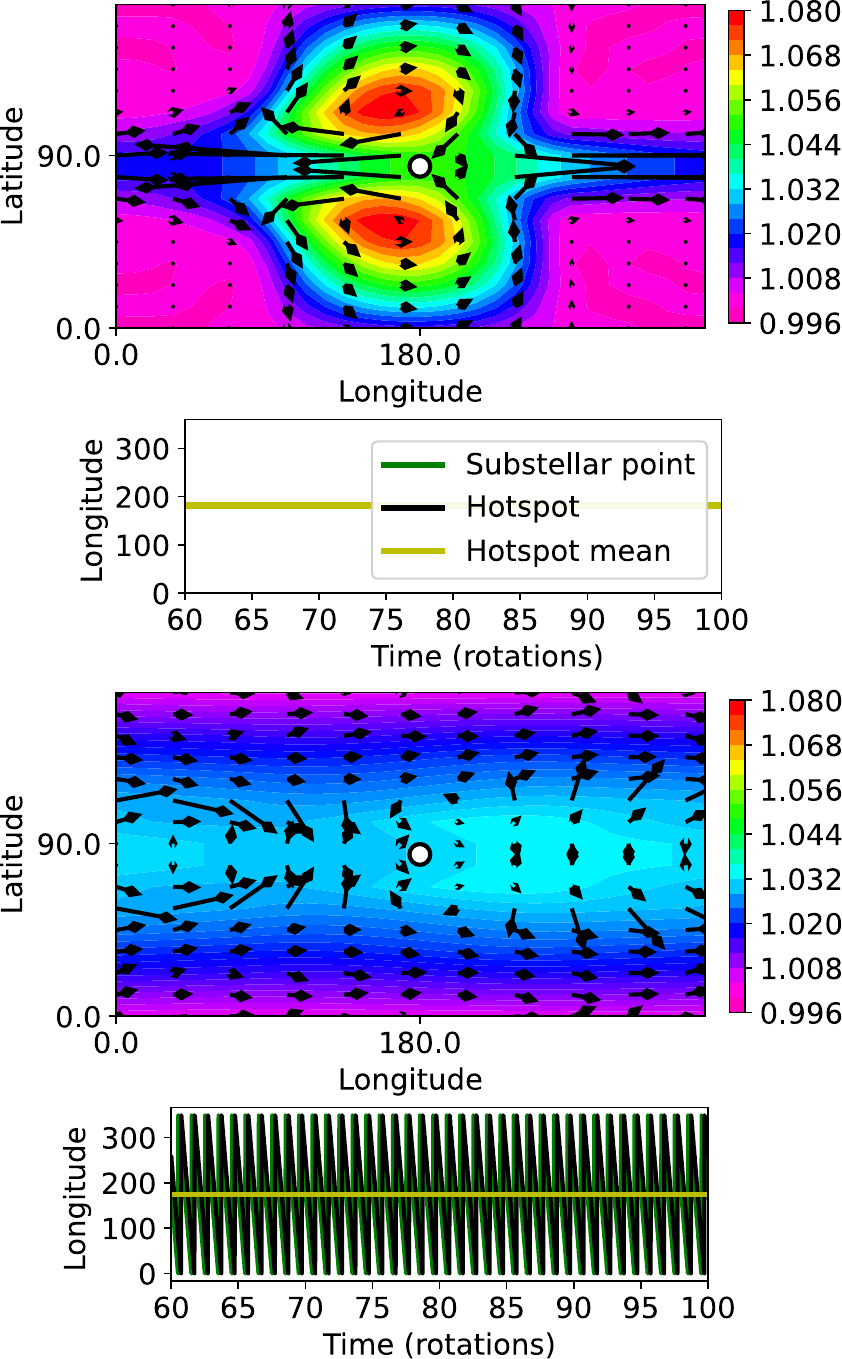}
    \caption{Top. Tidally synchronous circulation pattern for an Earth-like planet with a globally constant climate and the corresponding locations of the substellar point, the hotspot, and the latter's mean over 40 diurnal cycles. Bottom. The usual asynchronous circulation pattern for the same configuration with translating substellar point and hotspot. Note that the hotspot mean is the same in both cases i.e. at 180$^{\circ}$. For the asynchronous case, it is unintuitive to think of a mean longitude for the hotspot as it moves across the equatorial latitude, but we show it for comparison with the former and subsequent cases.}
    \label{fig:fig2}
\end{figure}

\section{Results}\label{sec:3}

\begin{figure*}
    \centering
\includegraphics[width=\linewidth]{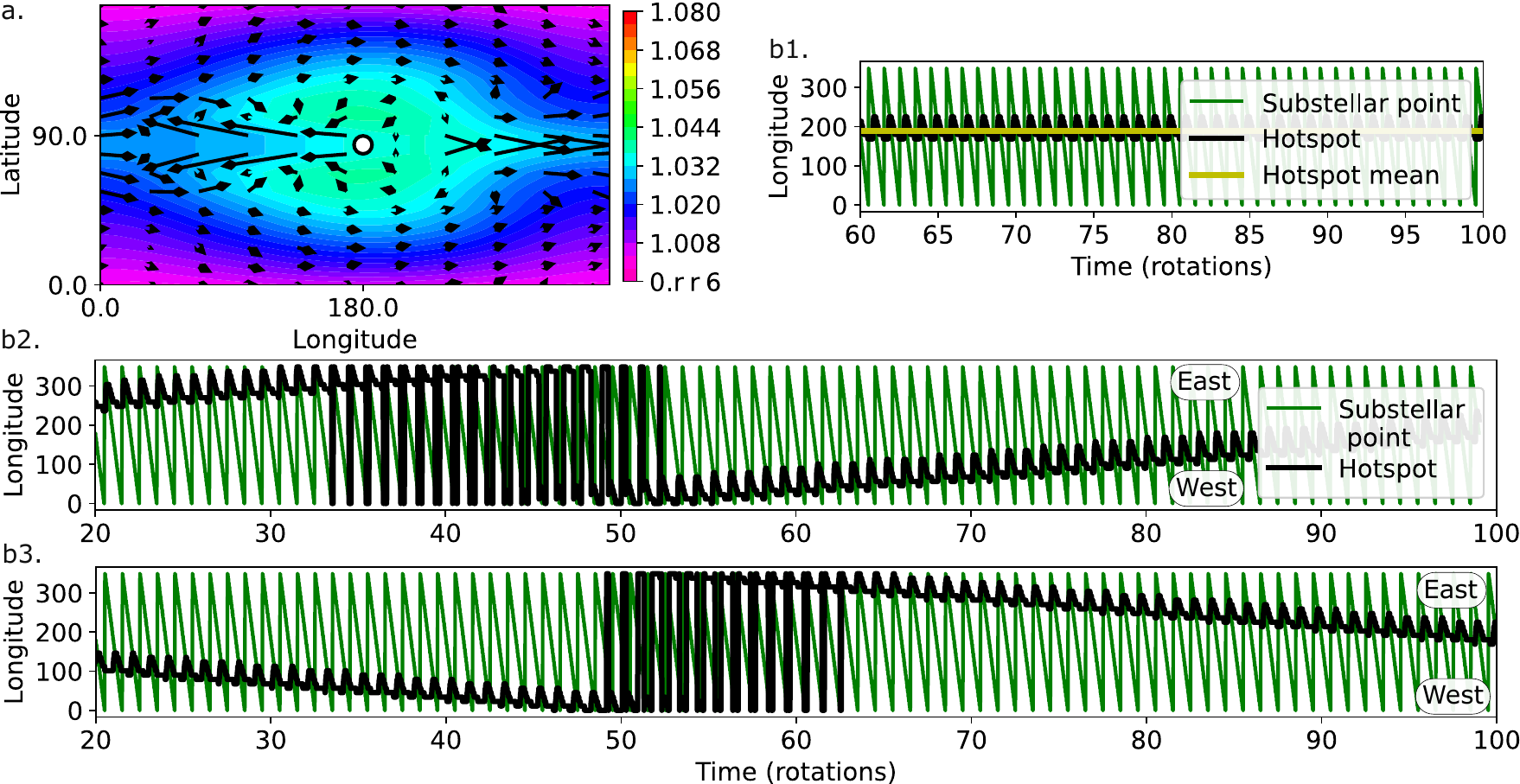}
    \caption{Beating planets mimic synchronous circulation patterns despite being asynchronous and show long-term variability. a. Height map of an Earth-like asynchronous planet when subjected to time-varying instellation of period same as that of rotation (= 1 day). Note that the fixing of the hotspot to the planet-frame-of-reference makes it similar to the tidally synchronous case in Fig. \ref{fig:fig2} (top). The colorbar used is the same as above. b1-b3. Motion of substellar point (green) and the corresponding hotspot (black) for $\omega = 2 \pi / (1, 1.01$, and  $0.99)$ days$^{-1}$, respectively. When instellation is constant the hotspot simply lags the substellar point as in Fig. \ref{fig:fig2} (bottom). However, due to varying instellation, here we observe a fixed, an eastward, and a westward creeping hotspot, respectively, for the same remaining parameters. Note that the large oscillations between 0$^\circ$ and 360$^\circ$ in b2-b3 occur because 0$^\circ$ and 360$^\circ$ are effectively the same longitudes on a spherical planet.}
    \label{fig:fig3}
\end{figure*}

Fig. \ref{fig:fig2} shows the height distribution and wind patterns on a tidally synchronous and an asynchronous Earth-like planet without varying instellation in the top and bottom panels, respectively. The first features stronger height contrasts between the constantly irradiated dayside and the dark nightside \citep{showman2011equatorial}, while the second has an averaged behavior because of the diurnal motion of the substellar point \citep{ohno2019atmospheres}. Additionally, the hotspot is fixed to the planet-frame-of-reference for the first case, while it follows the substellar point in the second \citep{penn2017thermal}. We shall use these cases as references in the following sections.

\subsection{Beating: fixed and creeping hotspots, therminators}

Now we vary the instellation for the asynchronous case. Fig. \ref{fig:fig3} panel b1 shows the locations of the substellar point and the corresponding hotspot when $\omega=\Omega=2\pi/1$ days$^{-1}$. Comparing their motion with an asynchronous planet with constant instellation (Fig. \ref{fig:fig2}, bottom) where the hotspot follows the substellar point, we see that despite being asynchronous, the hotspot in the variably irradiated case is almost fixed to a particular longitude, much like a tidally synchronous planet (Fig. \ref{fig:fig2}, top). Physically, this happens as the planet completes a rotation in the same time as it takes to complete a full cycle of incident stellar flux from a maxima to a minima back to a maxima. Opposite faces of the planet alternately experience the strongest and the weakest phases of irradiation, making one face perpetually hotter than the other. This is also evident in the circulation pattern in Fig. \ref{fig:fig3}a, which despite the milder height contrasts compared to the synchronous circulation, resembles it more than the asynchronous case. The same colorbar has been used for all three figures.

Fig. \ref{fig:fig3} panels b2 - b3 show the hotspot motion for $\omega$ slightly smaller ($\omega = 2\pi/1.01$ days$^{-1}$) and larger ($\omega = 2\pi/0.99$ days$^{-1}$) than the rotation frequency $\Omega$. For the former, the hotspot is seen to move from the west to the east with a period much longer than the rotation period. Contrarily, when $\omega$ is larger, the hotspot creeps from east to west, opposite to the direction of motion of the substellar point. This can be understood as the instellation peaking after or before the planet completes one rotation, respectively, shifting the height maxima set up by the instellation in the previous cycle slightly. The long-period movement of the hotspot is similar to the lower envelope frequency formed by the linear superposition of waves that form beats. Through the rest of the article, we shall collectively refer to the above as the `beating' phenomena. 

{\color{black}  It needs to be clarified here that the hotspot does not move with respect to the substellar point, but globally in the planet frame of reference. Whether the hotspot is east or west of the substellar point for an asynchronous planet is determined by the ratio of the speed of the substellar point with the gravity wave speed ($\sqrt{g\text{H}}$) for both prograde and retrograde substellar point motion as given in \cite{penn2017thermal}. In our simulations, the substellar point is always taken to be prograde i.e. both the planet's surface and the substellar point move in the same direction. For the control simulation (Table \ref{tab:contr}), the speed of the substellar point is 465 m/s and that of gravity waves is 31 m/s. From \cite{penn2017thermal}, this falls in the regime where the hotspot always lags the substellar point (see the $\alpha > 1$ regime in their Fig. 3). However, even if the gravity wave speed was to be greater than the speed of the substellar point, we expect the beating physics to still be consistent as it is independent of the location of the hotspot relative to the substellar point.
}

When $\omega=\Omega$, the planet-frame fixed hotspot indicates the creation of permanent local temperature extremes on opposite faces of the planet. The separating longitudes thus have relatively moderate temperatures. We call these thermal divides the `therminators', similar to `terminators' that separate the dayside from the nightside. In the case of tidally locked planets, the therminators mostly coincide with the terminators. Usually, even for asynchronous planets, the therminators would closely overlap with the terminators but move around the planet due to the diurnal motion of the substellar point. However, the syncing of diurnal and instellation frequencies fixes the former to the planet while the latter performs its regular motion. Since beating could essentially happen for any kind of planet with a sufficiently thick atmosphere, therminators of asynchronous CBPs may become important for habitability concerns, much like terminators of tidally locked planets \citep{lobo2023terminator}. This could effectively extend/shrink circumbinary habitable zones and is discussed later.

\newpage




\subsection{Period of variability}\label{sec:theo}

To support the phenomena analytically, the time-dependent part of the equilibrium height forcing in eq. (\ref{eq:4}) is broken down into individual cosine forcing functions as follows,
\begin{multline}
    (1 + \bar{\rm{f}} \cos \omega t)\cos\lambda_0(t) = \cos\lambda_0(t) \\+ \frac{\bar{\rm{f}}}{2}\cos(\lambda_0(t) + \omega t) + \frac{\bar{\rm{f}}}{2} \cos(\lambda_0(t) - \omega t). \label{eq:6}
\end{multline}
From eq. \eqref{eq:5} we have $\lambda_0(t) = \Omega t$. The argument of the last term on the RHS is thus $(\Omega-\omega)t$. {\color{black} Since both $\Omega$ and $\omega$ are positive, this corresponds to the magnitude-wise lowest-frequency component of the decomposition when the rotation and instellation variation rates are synced i.e.  $\Omega \sim \omega$. This component is responsible for the stationary/creeping behavior of the hotspot shown in Fig. \ref{fig:fig3}. The remaining components are related to the diurnal and smaller periods which can be seen as small localized motions of the hotspot in Fig. \ref{fig:fig3}b. Specifically, when $\lambda_0(t) = \omega t$, the hotspot is fixed to a particular longitude. When $\omega$ is slightly smaller or bigger than $\Omega$, we see the west-east and east-west motions of the hotspot with a long period given by $2 \pi / |\Omega - \omega|$ corresponding to the lowest-frequency component ($|\Omega - \omega|$).}

The decomposition of the varying instellation in eq. \eqref{eq:6} can be conceived as a linear superposition of three distinct sources of \textit{constant} radiation with the corresponding substellar points moving at speeds related to the component frequencies. For the same choice of parameters, we run individual simulations of the three substellar point motions with constant instellation and find that the slowest case (slowest component) shows maximum zonal height contrast given by the theoretical limit obtained in eq. (17) of \cite{perez2013atmospheric} that corresponds to a system strongly dominated by rotational effects. Indeed, the planetary Rossby number \citep{penn2017thermal} given by $\sqrt{g\text{H}}/(a \Omega)$ is $ 0.068$ for our system which is much less than 1, and hence corresponds to the Coriolis-dominated regime. For the remaining higher-frequency components of the decomposition, the zonal height contrast is found to be smaller than the slowest substellar point, likely because of the averaging role of the faster substellar point motion. Thus, the slowest component substellar point determines the net motion of the hotspot for the combined case of varying instellation. 

Note that the above justification is empirical and further systematic exploration beyond the parameter space explored in \cite{penn2017thermal} is required to understand the effect of substellar point motion on the atmospheric circulation of asynchronous planets.

\subsection{Dynamical regimes} \label{sec:dynreg}

For the rest, we take $\Omega=\omega$ to collectively address beating ($\Omega\sim\omega$). We also maintain $\tau_{\text{rad}}=\tau_{\text{drag}}$ for simplicity. All the shallow water simulation parameters (cf. eq. \ref{eq:1}-\ref{eq:5}) can be effectively studied using two dimensionless quantities that are known to control heat redistribution on planets known from previous work: $\tau_{\text{rad}}/$P$_\text{rot}$ and $\tau_{\text{wave}}/\sqrt{\tau_{\text{rad}}\text{P}_\text{rot}}$. The choice of the first parameter is motivated by the dynamical regimes of day-night temperature contrast and diurnal mean instellation for asynchronous planets \citep{ohno2019atmospheres}. The second demarcates the transition from a wave-adjusted uniform temperature distribution to a large day-night temperature contrast relative to radiative equilibrium for synchronous planets, applicable in the limit of weak drag (eq. 24, \citealp{perez2013atmospheric}). Here, $\tau_{\text{wave}}=L/\sqrt{g\text{H}}$ is the wave timescale or the time required for a barotropic disturbance to propagate over planetary scales, $a$ is the planetary radius, and $\sqrt{g\text{H}}$ is the speed of gravity waves. We take $L=L_{\text{equator}}=(a\sqrt{g\text{H}}/2\Omega)^{1/2}$, the equatorial Rossby deformation radius. This is a common assumption for barotropic atmospheric dynamics \citep{perez2013atmospheric}. The wave timescale has been shown to be crucial for the redistribution of heat in planetary atmospheres \citep{hammond2018wave} and will be seen to play a role in determining the regime of beating.

\begin{figure}
    \centering
\includegraphics[width=\linewidth]{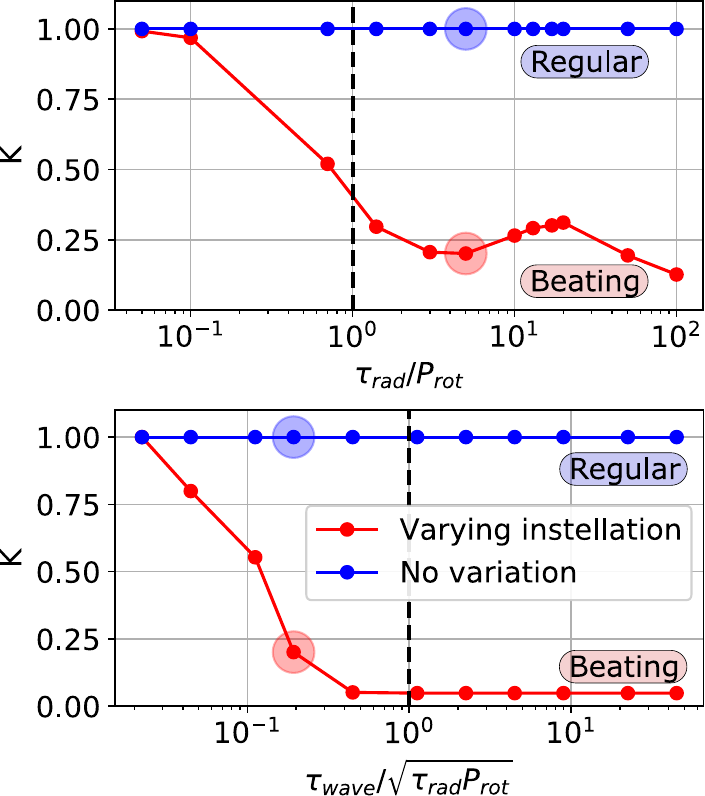}
    \caption{Plot of parameter K (normalized) varying with $\tau_{\text{rad}}/$P$_\text{rot}$ (top),  and $\tau_{\text{wave}}/\sqrt{\tau_{\text{rad}}\text{P}_\text{rot}}$ (bottom) for cases of varying and constant instellation. K is a metric measuring the degree of beating or climatic departure from the no-variation case. The vertical lines resemble the equality of relevant timescales. Large $\tau_\text{rad}$ and $\tau_\text{wave}$, and small P$_\text{rot}$ permit beating. {\color{black} The halo corresponds to the Earth-like control simulation presented before.}}
    \label{fig:fig4}
\end{figure}

To differentiate beating from the usual asynchronous no-variation case, we define a metric K, that measures the degree of latitudinal localization of the hotspot over the diurnal cycle. We obtain the mean longitude of the hotspot over multiple daily cycles from each simulation, and K is the standard deviation of this time-varying longitude from its mean. Or, K$=\sqrt{\sum (\lambda-\bar{\lambda})^2 / (N-1)}$, where $\lambda$ represents $N$ discrete values of the hotspot longitude recorded in time and $\bar{\lambda}$ is the mean. Thus, for a beating atmosphere, where the hotspot lingers close to the mean (Fig. \ref{fig:fig3} panel b1) i.e. is almost fixed, the K value is low, while for a hotspot that follows the substellar point (Fig. \ref{fig:fig2}, bottom), K is large. In general, a low-value K may be interpreted as a climatic departure of the VIP from its no-variation case. We now study the behavior of K based on the two ratios mentioned above.

Fig. \ref{fig:fig4} (top) plots the normalized value of K with $\tau_{\text{rad}}/$P$_\text{rot}$ for constant and varying instellation keeping $\tau_\text{wave}$ constant. {\color{black} In this figure, we have kept $\tau_{\text{rad}}$ constant and varied P$_\text{rot}$. However, we also checked the reverse for consistency. The transition to beating occurred at similar values of $\tau_{\text{rad}}/$P$_\text{rot}$ and the results showed a good qualitative match.} The Earth-like case, shown to feature beating in Fig. \ref{fig:fig3}, has a significant difference between K values for varying and constant instellation (halo around the corresponding point). This lies in the regime of diurnal mean instellation being the dominant form of forcing for asynchronous planets \citep{ohno2019atmospheres} which occurs when the radiative timescale is larger than the rotation period ($\tau_{\text{rad}}/$P$_\text{rot}=5$). As $\tau_{\text{rad}}/$P$_\text{rot}$ is reduced, we find varying instellation to have lesser effect on K. When $\tau_{\text{rad}}<$P$_\text{rot}$, the day-night forcing pattern determines the temperature map of the immediately irradiated region, making the hotspot closely follow the substellar point, thus increasing K and preventing beating. 

In Fig. \ref{fig:fig4} (bottom) we probe the role of $\tau_\text{wave}$ keeping $\tau_{\text{rad}}$ and P$_\text{rot}$ constant. {\color{black} Here, we vary $\tau_\text{wave}$ from the Earth-like case (halo, $\tau_{\text{wave}}/\sqrt{\tau_{\text{rad}}\text{P}_\text{rot}}=0.1926$) by either changing the scale height H or the radius of the planet R, yielding similar results as far as the transition to beating is concerned.} As $\tau_{\text{wave}}/\sqrt{\tau_{\text{rad}}\text{P}_\text{rot}}$ increases, the planet switches to beating in the presence of synced varying instellation. From \cite{perez2013atmospheric} it is known that when $\tau_{\text{wave}}<<\sqrt{\tau_{\text{rad}}\text{P}_\text{rot}}$, the steady-state planetary circulation of synchronous planets exhibits minimal height variation, deviating strongly from the forced day-night radiative equilibrium field. The same condition appears to apply here for the limit of no-beating. Consequently, the regime of stronger height contrasts akin to the radiative equilibrium field marked by longer wave timescales shows the occurrence of beating. 

\begin{figure*}
    \centering
\includegraphics[width=\linewidth]{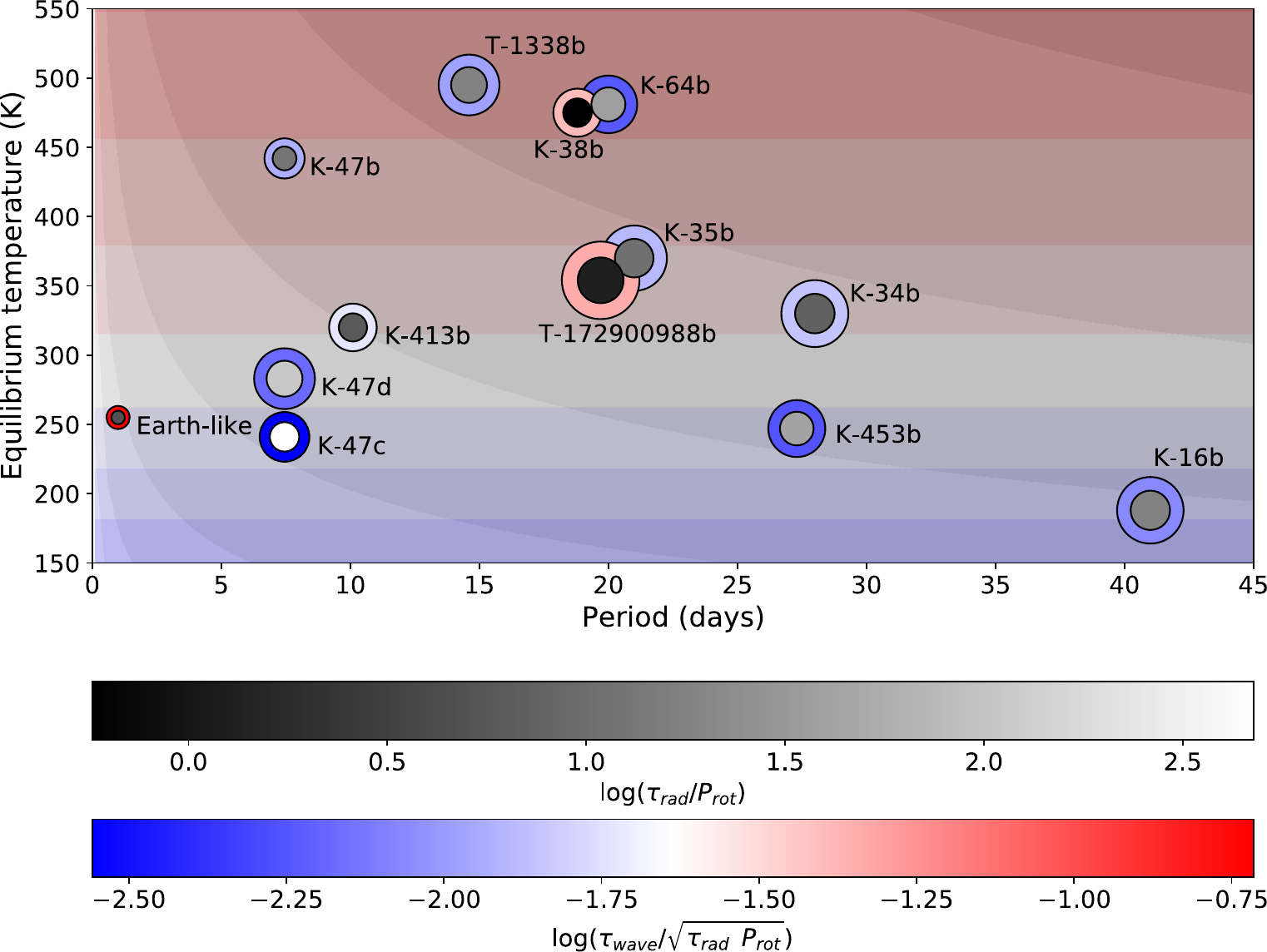}
    \caption{$\tau_{\text{rad}}/$P$_\text{rot}$  and $\tau_{\text{wave}}/\sqrt{\tau_{\text{rad}}\text{P}_\text{rot}}$ values for different Kepler and TESS circumbinary planets. The planets are scaled by their radius. The background contour map is specifically set to properties similar to Earth-like planets. The colorbars represent the planets quantitatively and the background qualitatively. In general, climatic departures are likely where the colors are brighter (both red and white) implying that faster-spinning, hotter planets (top-left corner) are more susceptible to beating. }
    \label{fig:fig5}
\end{figure*}

Explaining the phenomena more pedagogically, beating/climatic departure requires the atmosphere to have memory of previous heating, such that the existing temperature distribution can be strengthened in the next cycle. A large radiative cooling time $\tau_{\text{rad}}$ is helpful by definition. Additionally, fast rotation, aids beating by sustaining flow structures created by radiative forcing via geostrophic balance \citep{perez2013atmospheric}. Both the above points to the right of the dashed vertical line in Fig. \ref{fig:fig4} (top) where $\tau_{\text{rad}}/$P$_\text{rot}>1$. However, if $\tau_{\text{wave}}$ is too small, wave adjustment processes may redistribute the built-up temperature pattern fast, leading to a loss of memory (left of dashed vertical line in Fig. \ref{fig:fig4}, bottom). Alternately, it may be hypothesized that if drag were to be strong, it could potentially prevent wave adjustment from redistributing heat as much, further contributing to memory. The role of drag is briefly explored in appendix \ref{app:B} for the case of CBP Kepler-47b. 

While the above are accurate dynamical arguments explaining the phenomena at first-order, the problem warrants further exploration, possibly using linearized theories of the shallow water system to analytically establish the conclusions made above. We have left that for future work. 

\section{Discussion}\label{sec:4}

In this section, we explore the possibility of observed CBPs to exhibit climatic departures based on the criteria described in the previous section. We also expand on the habitability characteristics of CBPs provided beating occurred.

\subsection{Applicability to real systems}

We calculate $\tau_{\text{rad}}/$P$_\text{rot}$  and $\tau_{\text{wave}}/\sqrt{\tau_{\text{rad}}\text{P}_\text{rot}}$ values for the set of Kepler and TESS circumbinary planets and plot them in the T$_\text{eq}$---P$_\text{rot}$ space in Fig. \ref{fig:fig5}. The x-axis represents both the period of instellation variation due to the binary orbit P$_\text{bin}$, and the rotational period P$_\text{rot}$, which are assumed equal. This is a strong assumption but justified by the lack of observational constraints on planetary spins in general. The eccentricity of planetary orbits is neglected as it is expected to induce variability in timescales similar to the orbital period which is at least 5 times larger than the assumed rotation period of all planets. Even if this assumption was relaxed to account for the contribution of orbital motion to the diurnal period of a planet, the flexibility on rotation rate allows us to adjust it in such a way that the binary period equals the diurnal period of the planet (cf. eq. 4, \cite{penn2017thermal}), the condition necessary for beating. 

Based on the two criteria established in \S \ref{sec:dynreg}, the CBPs shown in brighter colors must belong to dynamical regimes supporting climatic departures from the no-variation case. For most CBPs, $\tau_{\text{rad}} \gtrsim $P$_\text{rot}$ (see Table \ref{tab:planetary_data_ratios}) satisfying the first criteria (grey/white on colorbar). The outlier Kepler-38b, for which $\tau_{\text{rad}} / $P$_\text{rot} \approx 0.57$, still makes the cut for climatic-departure according to Fig. \ref{fig:fig4} (top) even though it may not qualify for the beating extreme. However, most CBPs also have $\tau_{\text{wave}}<<\sqrt{\tau_{\text{rad}}\text{P}_\text{rot}}$ (blue on colorbar) making them less likely candidates, except for Kepler-38b, TIC-172900988b, and the Earth-like CBP. Thus, if at all, these three planets should be the most suitable for climatic departures (redder in Fig. \ref{fig:fig5}). We already know from Fig. \ref{fig:fig3} that the Earth equivalent indeed qualifies. Additionally, to confirm our hypothesis, we simulate each planet individually. We find only these three planets to have K values noticeably lower than the corresponding no-variation, case confirming our hypothesis. Further, none of the planets except for the circumbinary Earth, which has the largest value of $\tau_{\text{wave}}/\sqrt{\tau_{\text{rad}}\text{P}_\text{rot}}$ among all, exhibit beating. This is because, despite satisfying the criterion of large $\tau_\text{rad}$, all planets have a tiny $\tau_\text{wave}$ which allows for rapid redistribution of heat, aiding heat-memory loss and preventing climatic departures. The planet properties and details of calculation of relevant quantities are given in appendix \ref{app:A}.

The background in the Fig. \ref{fig:fig5} shows the contours of $\tau_{\text{rad}}/$P$_\text{rot}$  and $\tau_{\text{wave}}/\sqrt{\tau_{\text{rad}}\text{P}_\text{rot}}$ for Earth-like planets as T$_\text{eq}$ and P$_\text{rot}$ are varied. This means that except for T$_\text{eq}$ and P$_\text{rot}$, every other planetary parameter is assumed to be Earth-like. Based on the first criterion, climatic departure is more likely for planets with shorter periods and lower equilibrium temperatures i.e. the bottom left corner of the plot. The second criterion, is found to be independent of P$_\text{rot}$ for our choice of planetary length scale $L=L_\text{equator}$ (cf. \S \ref{sec:dynreg}), indicating that higher T$_\text{eq}$ is the sole driver of climatic departures. Combining both the above, we conclude that hotter and faster spinning planets (top left corner) are best suited for climatic departures, where both $\tau_{\text{rad}}/$P$_\text{rot}$  and $\tau_{\text{wave}}/\sqrt{\tau_{\text{rad}}\text{P}_\text{rot}}$ are large. However, in addition to the rotation and instellation variation periods, the host star type, atmospheric composition of the planet, and its semi-major axis, all contribute to the aforementioned quantities, making the identification of the exact region of parameter space more involved. This may be further complicated by other physical processes including realistic radiative transfer, clouds and water-vapor feedback, and hence require detailed 3D climate simulations to map the regime of beating.

\subsection{Consequences on habitability}


The fixing/creeping of the hotspot/therminators to the planet during beating may have interesting implications for habitability. A beating planet inside the traditional circumbinary habitable zone (green shade, ref. Fig. \ref{fig:fig1}a) will have reduced fractional habitability relative to the no-variation case because of temperature extremes on the hot and cold sides. However, the therminators may be temperate. Recent studies have shown that water-limited tidally locked M-dwarf rocky planets are great candidates for terminator habitability \citep{lobo2023terminator, lobo2024climate}. Even though our model is relatively simplistic, the prospect of habitable therminators is a natural extension of the earlier work and requires further investigation. Interestingly, because of the diurnal cycle of asynchronous planets, the permanent/creeping therminator still receives sunlight and darkness alternatively, something crucial for the development of photosynthetic capacity \citep{Caraveo2021}, unavailable for tidally locked planets. Additionally, the creeping hotspot cases show longer-than-diurnal climate variability, akin to mild Exo-Milankovich cycles \citep{deitrick2018exo}. For all these reasons, beating planets may be considered intriguing candidates for habitability. 

Figure \ref{fig:fig1}a also shows optimistic extensions of the circumbinary habitable zone created by beating planets. Such planets within the inner edge (blue shade) may feature a reasonably warm coldside, while one beyond the outer edge (yellow shade) may have a reasonably cool hotside, both suitable for habitability. However, the caveat is that our model does not include nuances like water vapor and cloud physics, which are known to significantly impact heat redistribution via latent heat and albedo. Research suggests that inner edge tidally-locked planets risk runaway greenhouse scenarios making them uninhabitable \citep{kopparapu2018habitable}, while the outer edge of HZs may likely be extended by instellation variation from eccentricity \citep{dressing2010habitable}. Our model is relatively simplistic, and hence the cumulative effect of different contributions to instellation variation will need further GCM simulations to hone in on the question of habitability. 

\subsection{Model limitations}\label{sec:lim}
Our shallow-water equations assume a thin fluid layer, with a horizontal scale much larger than the vertical, an assumption unsuitable for thicker atmospheres. It omits the representation of vertical structure/transport and baroclinic dynamics present in 3D climate models, affecting planetary heat redistribution \citep{penn2017thermal}. It also lacks detailed radiative transfer, frictional diffusivity, and compositional diversity. For smaller terrestrial planets, other physics like land mass distribution, clouds, water vapor, and ice-albedo feedback may be important but are beyond the scope of the current model. Additionally, we use a simple periodic forcing term for stellar insolation which may have complicated variability patterns \citep{may2016examining}. Moreover, the fractional semiamplitude $\bar{\text{f}}$, has been restricted to a constant value for this work. 

{\color{black} We have also assumed our planet to have zero eccentricity and obliquity. While eccentricity may be modeled as a periodic variation of instellation, keeping the current analysis valid to some extent, a highly oblique planet may show features like near-polar hotspots \citep{rauscher2017models} and significantly different atmospheric circulation \citep{lobo2020atmospheric} under which synchronous and asynchronous rotation, crucial for this study, may not be defined. However, we expect small obliquity asynchronous planets, for which seasonal variations are mild, to still exhibit beating. Nonetheless, the detailed study of diversely oblique planets under variable instellation warrants further research.} 

Despite the above caveats, our 2D model captures large-scale weather phenomena on a sphere which has been demonstrated to be primarily an interaction of radiative, rotational, drag, and wave-adjustment processes \citep{showman2011equatorial,showman2012doppler,hammond2018wave}, features that collectively explain the mechanism behind beating. Aside from incorporating the aforementioned physics, further extensions using the shallow water model may involve varying the irradiation frequency markedly beyond the diurnal frequency to twice or half its value. Here, we have restricted ourselves only to beating and left the rest to future work.

\subsection{Phase curves of beating planets}

{\color{black} The enhanced contrast caused by a fixed hotspot could lead to noticeable phase-curve variations, usually unlikely for asynchronous planets at significant distances from the host star (large $\tau_{\text{rad}}$). For such planets, the period at which the phase curve peaks would be the same as the rotation period of the planet. If the hotspot creeps, the phase-curve period would change slightly due to the additional motion of the hotspot. Since the period of such motion is known from \S \ref{sec:3}, the period of the phase curve may also be determined. We may even obtain synthetic phase curves from the shallow water model following \cite{penn2017thermal} and verify our hypothesis. However, given our model limitations, a detailed discussion on observational consequences would be premature and hence will be addressed in a later paper.}

\section{Conclusion} \label{sec:5}

In this paper we have explored the atmospheric circulation of non-eccentric-tilted, asynchronous, thick atmosphere planets on circular orbits for which variations in the instellation have a non-negligible impact on planetary circulation. This is contrary/complementary to the parameter space probed in \cite{may2016examining} and \cite{popp2017climate} for circumbinary planets where diurnal averaging allows the use of latitudinal energy balance models, leading to close similarities with the equivalent single star (no-variation) case. Using our instellation-source-agnostic approach, we simulate planetary atmospheres with a 2D two-layer shallow water model and point out regimes of circulation that vary significantly from the corresponding constant irradiation scenarios.

Specifically, we find that the atmospheric circulation of asynchronous planets may resemble that of tidally synchronous ones when the instellation frequency is equal to the diurnal frequency, leading to a hotspot that is stationary in the planet's frame of reference. For minor deviations from the diurnal frequency, the hotspot is observed to creep slowly across the surface of the planet at a rate much slower than the motion of the substellar point, akin to the lower envelope frequency of two signals forming a beat pattern. This phenomenon is named `beating', and the rate of hotspot motion is found to be the difference between the planet's diurnal and instellation variation frequencies. We provide a first-order analytical explanation for this phenomenon by decomposing the variable stellar forcing into constant individual components that describe the overall climate of the planet.

Further, we delineate the dynamical space within which climatic departure/beating occurs by referencing two important timescale ratios from literature: $\tau_{\text{rad}}/$P$_\text{rot}$ \citep{ohno2019atmospheres}  and $\tau_{\text{wave}}/\sqrt{\tau_{\text{rad}}\text{P}_\text{rot}}$ \citep{perez2013atmospheric}. They demarcate regimes of strong day-night contrast from averaged flows driven by radiative/Coriolis and wave physics, respectively. We find that the planetary circulation is independent of the style of radiative forcing (constant or periodic) in the limit of slow rotation, strong radiation, and fast wave-adjustment processes. Otherwise, climatic departures may be observed. Based on this distinction, we classify different Kepler and TESS circumbinary planets into candidates likely or unlikely to show climatic departures and find simulations to confirm our classification, establishing the classifying criteria. Considering the possibility of detected CBPs to exhibit climatic departures, and a hypothetical Earth-like CBP being the strongest candidate for beating, we discuss its implications on circumbinary habitable zones: we project the creation of habitable hot-cold divides, or `therminators', similar to terminator habitability \citep{lobo2023terminator}, and optimistic extensions of existing habitable zones.

Future work includes exploring this specific region of parameter space with 3D global climate models that account for radiative transfer, water-vapor/ice-albedo feedback, clouds, and other physical nuances for Earth-like CBPs. The impact of beating on planetary light curves would be another direction of investigation. The dynamical limits of beating, including drag, can be explored with linearized shallow-water theory to uncover the underlying mechanisms. It is also crucial to explore the probable causes, for the synchronization of diurnal and instellation variation frequencies, potentially involving thermal tides \citep{leconte2015asynchronous}.

\section*{Acknowledgements}
The author is grateful to Thaddeus David Komacek and Marta L. Bryan for providing valuable feedback on an early draft of the paper improving it significantly. The author thanks the anonymous reviewers for their constructive comments. The author thanks Hayley Q. Beltz, Akhilesh S. Tiwari, Abdulmuttalib Lokhandwala, Mykhaylo Plotnykov, Garett Brown, Eesha Dasgupta, Geoffrey K. Vallis, Morgan O'Neill, and Kristen Menou, for inputs through discussions. The author also thanks David V. Martin for sharing a list of TESS circumbinary planets. Finally, the author thanks Arpita Banik and Ranjit Banik for their constant encouragement. 

The author is supported by the Department of Physics, University of Toronto, and the computations were performed using the Niagara supercomputer operated by SciNet under the Digital Research Alliance of Canada.

\vspace{5mm}


\software{numpy \citep{harris2020array}, Dedalus3 \citep{burns2020dedalus}}
\appendix

\begin{table}
\centering
\begin{tabular}{cccccccccc}
\hline
Planet & Mass & Radius & T$_\text{eq}(K)$ & P$_\text{bin}$ & $\tau_\text{wave}$ & $\tau_\text{rad}$ & $\frac{\tau_{\text{rad}}}{\text{P}_\text{rot}}$ & $\frac{\tau_\text{wave}}{\sqrt{\tau_{\text{rad}}\text{P}_\text{rot}}}$ & References \vspace{1mm}\\ \hline \hline
Earth-like & 1 & 1 & 255 & 1 & 0.4308 & 5 & 5 & 0.1926 & \cite{penn2017thermal}\\
K-16b & 105 & 8.3 & 188 & 41 & 1.5031 & 731 & 17.8 & 0.0087 & \cite{kane2012habitable, welsh2012recent} \\
K-34b & 70 & 8.4 & 330 & 28 & 1.0863 & 208 & 7.4 & 0.0142 & \cite{kane2012habitable, welsh2012recent} \\
K-35b & 40 & 7.9 & 370 & 21 & 0.8925 & 232 & 11.06 & 0.0128 & \cite{kane2012habitable, welsh2012recent}\\
K-47b & 8.4 & 3.03 & 442 & 7.45 & 0.3132 & 94 & 12.6 & 0.0118 & \cite{welsh2012recent}\\
K-47c & 3.2 & 4.62 & 241 & 7.45 & 0.4500 & 3550 & 476.5 & 0.0028 & \cite{welsh2012recent,orosz2019discovery} \\
K-47d & 19 & 6.9 & 283 & 7.45 & 0.5288 & 825 & 110.8 & 0.0067 & \href{https://www.nasa.gov/missions/kepler/discovery-alert-a-third-planet-in-kepler-47-system/#:~:text=Kepler-47d%E2%80%99s%20equilibrium%20temperature%20is%20roughly%2050%20degrees%20F,Kepler-47c%20is%20%E2%80%9126%20degrees%20F%20%28%E2%80%9132%20degrees%20C%29.}{www.nasa.gov} \\
K-64b & 3.38 & 6.2 & 481 & 20 & 0.7187 & 761 & 38.06 & 0.0058 & \cite{schwamb2013planet}; Planet Hunters \\
K-413b & 67 & 4.25 & 320 & 10 & 0.4685 & 61.4 & 6.07 & 0.0188 & \cite{kostov2014kepler} \\
K-453b & 16 & 6.06 & 247 & 27.3 & 0.9811 & 1137 & 41.6 & 0.0056 & \cite{welsh2015kepler} \\
K-38b & 122 & 4.35 & 475 & 18.8 & 0.5855 & 10.7 & 0.57 & 0.0411 & \cite{orosz2012neptune} \\
T-1338b & 11.3 & 6.9 & 495 & 14.6 & 0.6431 & 258 & 17.7 & 0.0105 & \cite{kostov2020toi,Wang_2024} \\
T-172900988b & 824 & 11.24 & 354 & 19.7 & 1.0369 & 25.7 & 1.3 & 0.0460 & \cite{kostov2021tic} \\ 
\end{tabular}
\caption{Kepler and TESS circumbinary planet parameters. Mass and radius data has been acquired from the NASA Exoplanet archive (\href{https://exoplanetarchive.ipac.caltech.edu/}{https://exoplanetarchive.ipac.caltech.edu}). The specific references are for the equilibrium temperature.}
\label{tab:planetary_data_ratios}
\end{table}
\begin{figure*}
    \centering
\includegraphics[width=0.8\linewidth]{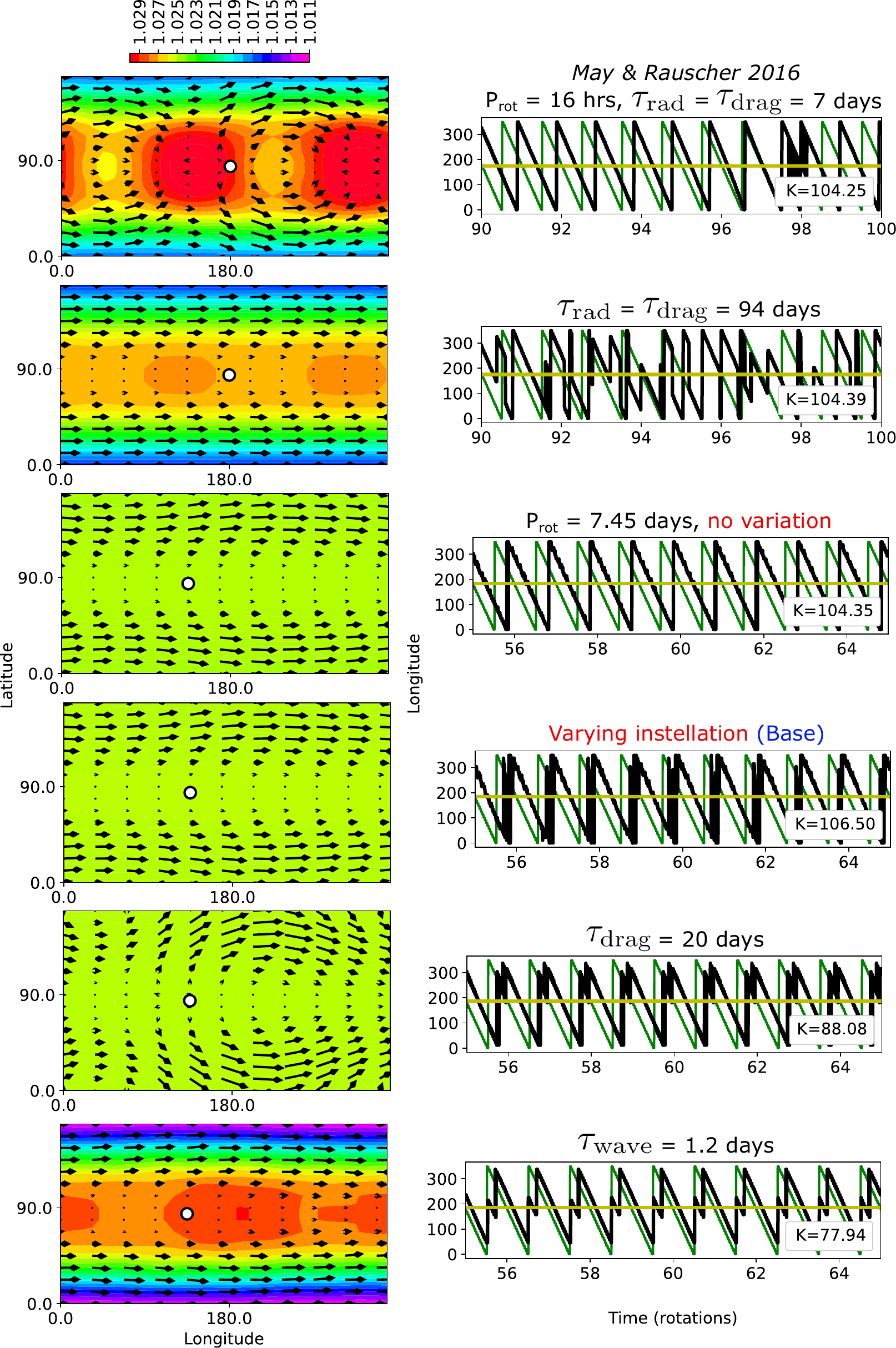}
    \caption{Atmospheric circulation patterns and corresponding hotspot motions for different cases of Kepler-47b. Each panel displays a simulation with parameters consistent with the one above, differing only in the quantities specified in the respective heading. We see a reduction of K magnitude and hotspot jumps/resets in the cases of strong drag (panel 5) and when $\tau_\text{wave}$ is significantly increased (panel 6). Thus, drag and wave physics are important players in climatic departures.}
    \label{fig:fig6}
\end{figure*}

\section{Climatic departures on Kepler circumbinary planets} \label{app:A}
We calculate the value of $\tau_{\text{rad}}/$P$_\text{rot}$  and $\tau_{\text{wave}}/\sqrt{\tau_{\text{rad}}\text{P}_\text{rot}}$ (cf. \S \ref{sec:dynreg}) for Kepler and TESS circumbinary planets. We use the formulae below, following from the descriptions given in \cite{perez2013atmospheric} and \cite{landgren2023shallow}:
\begin{equation}
    \tau_{\text{wave}} = \frac{L_{\text{equator}}}{\sqrt{g\text{H}}}, \\
    L_{\text{equator}} = \sqrt{\frac{a\sqrt{g\text{H}}}{2\Omega}}, \\
    g = \frac{\mathcal{G} \text{M}}{a^2}, \\
    \text{H} = \frac{\mathcal{R}_{\text{gas}} \text{T}_{\text{eq}}}{g}, \mathcal{R}_{\text{gas}}=\frac{\mathcal{R}}{\mu_{\text{gas}}},  \text{ and }\\
    \tau_{\text{rad}} = \frac{\Delta p}{g} \frac{c_p}{4 \sigma \text{T}_{\text{eq}}^3}.
    \label{eq:A1}
\end{equation}
Here, $L_{\text{equator}}$ is the equatorial Rossby radius of deformation, $\sqrt{g\text{H}}$ is the speed of gravity waves, H is the scale height, $g$ is the surface acceleration due to gravity, $\mathcal{G}$ is the universal gravitational constant, $\mathcal{R}_{\text{gas}}$ is the specific gas constant, $\mu_{\text{gas}}$ is the mean molecular weight of the atmosphere, M, $\Omega$ and $a$ are the mass, rotation rate and radius of the planet, T$_{\text{eq}}$ is the equilibrium temperature, $\Delta p$ is the thickness of the weather layer, $c_p$ is the specific heat of the atmosphere at constant pressure, and $\sigma$ is the Stefan-Boltzman constant. For most planets, the M and $a$ are known from observations. T$_{\text{eq}}$ is a derived parameter available for most of the candidates from the corresponding references in Table \ref{tab:planetary_data_ratios}, except for TIC-172900988b \citep{kostov2021tic}. We make a crude approximation for the planet using the formula T$_{\text{eq}}=\text{T}^* (\sqrt{\text{R}^*/s})0.25^{0.25}$ \citep{kempton2018framework}, where T$^*(\approx 6000$K) and R$^*(\approx 1.3$R$_\odot)$ are the temperature and radius of the binary pair (stars are similar), and $s(\approx 0.867$au) is the semi-major axis of the CBP. Note here again the rotation rate of the CBP is assumed to be equal to the binary period for beating to occur. Thus, $\tau_{\text{rad}}/$P$_\text{rot}$  and $\tau_{\text{wave}}/\sqrt{\tau_{\text{rad}}\text{P}_\text{rot}}$ may be calculated from the following six quantities: M, $a$, T$_{\text{eq}}$, $c_p$, $\Delta p$, and $\Omega$. 

For all the planets here, we take $\Delta p$ to be approximately equivalent to the photosphere depth for Neptune-like planets $\sim 10^5$Pa. The atmospheres are assumed to be hydrogen dominated, with $c_p=14.304$ kJ/K-kg, and $\mu_\text{gas}=2.015\times 10^{-3}$kg/mol.

\section{The case of Kepler-47b}\label{app:B}

Kepler-47b is a circumbinary planet, well studied using energy balance models and global climate models \citep{may2016examining}. We first simulate the planet with the default parameter set ($\tau_\text{rad}$ = 7 days, P$_\text{rot} = 16$ hours, rest in Table \ref{tab:planetary_data_ratios}) of \cite{may2016examining} using our shallow water model. The value for P$_\text{rot}$ is set based on the similarity of the planet with Neptune, which has a rotation period of 16 hours. We find the atmospheric temperature distribution and wind pattern to qualitatively differ from their GCM results (Fig. \ref{fig:fig6} panel 1). This is most likely due to our assumption of $\tau_\text{drag}=\tau_\text{rad}$ making drag significantly strong. However, our $\tau_\text{rad}$ for Kepler-47b calculated from planet properties in Table \ref{tab:planetary_data_ratios} is larger, at 94 days. The second panel in Fig. \ref{fig:fig6} shows this low drag circulation which now matches \cite{may2016examining} more closely. We use these settings for further exploration.

Syncing rotation and instellation variation periods to 7.45 days (panel 4) does not cause significant climatic-departure i.e.  the value of K (cf. \S \ref{sec:dynreg}) remains the same as the no-variation (panel 3). In fact, the height contrasts reduce significantly due to the slower rotation rate. We refer to panel 4 as the base case. In panel 5 we probe the effect of increased drag by reducing $\tau_\text{drag}$ to 20 days. Alternatively, from the base case, we change the radius and mass of the planet to 50 R$_\oplus$ and $2287$M$_\oplus$, respectively, keeping gravitational acceleration and hence $\tau_\text{rad}$ the same (cf. eq. \ref{eq:A1}), but increasing the $\tau_\text{wave}$ to 1.2 days. This increases $\tau_\text{wave}/\sqrt{\tau_{\text{rad}}\text{P}_\text{rot}}$ to a value 0.048 comparable to Kepler-38b and TIC-172900988b (see Table \ref{tab:planetary_data_ratios}). 

There is a notable reduction in the value of K for both cases, implying climatic-departure. The time-dependent hotspot location does not show localization as happens for beating, but experiences sudden jumps/resets (Fig. \ref{fig:fig6} panels 5 and 6). However, the height contrasts in the strongly dragged case are much lower than when the wave timescale is larger. This reiterates the role of drag and wave physics in determining the nature and extent of climate variations of variably irradiated planets and establishes the need for more accurate resolution of physical processes through 3D GCM simulations.

\bibliography{sample631}{}
\bibliographystyle{aasjournal}



\end{document}